\documentstyle[12pt]{article}

\newcommand {\eqref} [1] {(\ref {#1})}
\newcommand {\beq} {\begin{equation}} 
\newcommand {\eeq} {\end{equation}}
 \newcommand {\ber}{\begin{eqnarray*}}
 \newcommand {\eer} {\end{eqnarray*}}
\newcommand {\bea}{\begin{eqnarray}}
 \newcommand {\eea} {\end{eqnarray}} 

\newcommand{\Nfour} {${\cal N}=4\ $}
\newcommand{\Ntwo}{${\cal N}=2\ $}
\newcommand{\None}{${\cal N}=1\ $}

\def\Acknowledgements{\bigskip  \bigskip {\begin{center} \begin{large}
             \bf ACKNOWLEDGEMENTS \end{large}\end{center}}}

\begin{document}\begin{titlepage}
\rightline{S742.1099}
\rightline{\today}
\vskip 1cm
\centerline{{\Large \bf A Note on Non-Commutative Orbifold Field
    Theories }}
\vskip 0.5cm
\centerline{Adi Armoni}
\vskip 0.5cm
\centerline{Centre de Physique Th{\'e}orique de l'{\'E}cole Polytechnique}
\centerline{91128 Palaiseau Cedex, France}
\vskip 0.5cm
\centerline{armoni@cpht.polytechnique.fr}
\vskip 0.5cm
\begin{abstract}
We suggest that orbifold field theories which are obtained from
 non-commutative \Nfour SYM are UV finite. In particular,
 non-supersymmetric orbifold truncations
might be finite even at finite values of $N_c$.
\end{abstract}
\end{titlepage}
\section{Introduction}
Recently non-commutative Yang-Mills theory \cite{con} attracted a lot of
attention, mainly due to discoveries of new connections to
 string theory\cite{CDS,DH,SW}.
In a recent paper\cite{BS}, it was suggested that
the divergences of the non-commutative Yang-Mills theory are dictated by the 
large $N_c$ limit of the theory. Namely, that divergences occur
only in planar diagrams (the observation that the planar commutative
and non-commutative theories are the same up to phases in the Green
 functions, was
already made in\cite{filk}. A careful analysis of the divergences was
 carried out in ref.\cite{KW}). 

Another direction of research is the study of orbifold field
theories - motivated by the AdS/CFT correspondence \cite{Mal}. 
It was conjectured
by Kachru and Silverstein \cite{KS} that orbifolds of $AdS_5 \times S^5$ which
 act on the $S^5$ part define a large $N_c$ finite theory, even when 
the R-symmetry is completely broken and the theory is not supersymmetric.
This conjecture was later proved using both field theory \cite{BJ} and string theory \cite{BKV} techniques.

In this note we would like to consider non-commutative Yang-Mills
 theories which are obtained by an orbifold truncation of \Nfour SYM.
We suggest that these theories are UV finite, namely that there are no 
divergent Feynman diagrams, even when the theory under consideration is
 non-supersymmetric and the number of colors is finite. Note, however,
 that our conjecture relies on recent suggestions \cite{BS,MSV} about
 UV finiteness of non-planar graphs in the non-commutative theory.
 The later were not fully proved yet, but seems to be necessary
if the non-commutative theory is renormalizable.

\section{Orbifold field theories}
Orbifold field theories are obtained by a certain truncation of
a supersymmetric Yang-Mills theory. Let us consider the special case
of \Nfour. The truncation procedure is as follows: consider a discrete
subgroup $\Gamma$ of the \Nfour R-symmetry group $SU(4)$. For each
element of the orbifold group, a representation $\gamma$ inside
$SU(|\Gamma |N_c)$  should be specified. Each field $\Phi$ transform
as $\Phi \rightarrow r \gamma ^\dagger \Phi \gamma$, where $r$ is a
 representation matrix inside the 
R-symmetry group. The truncation is achieved by keeping invariant fields.
The resulting theory has a reduced amount of supersymmetry, or no
supersymmetry at all. It was conjectured\cite{KS}, based on the AdS/CFT
 conjecture, that the truncated large $N_c$ theories are finite as the parent
\Nfour theory. Later it was proved \cite{BJ} that the planar 
diagrams of the truncated theory and parent theories are identical.
In particular it means that the perturbative beta function of the
large $N_c$ daughter theory is zero and that the theory is finite.

In the cases of \Ntwo truncations, there is only one-loop
 (perturbative) contribution to the beta function. Its
vanishing indicates the perturbative finiteness of the daughter theory at
finite $N_c$ as well.

In \None truncations the situation is more subtle. Indeed, the theory
is finite at finite values of $N_c$, but the finiteness is due to
Leigh-Strassler type of arguments\cite{LS}. In that case one should
consider the $SU(N_c)$ version of the theory (and not the $U(N_c)$
theory which is obtained from the string theory orbifold),
 since the $U(1)$ beta
function is always positive at the origin. In addition ${1\over
  N_c^2}$ shifts of the Yukawa couplings are needed\cite{OT}.

In the non-supersymmetric case there are no known examples of finite theories
at finite $N_c$, though attempts in this direction using orbifolds
were recently made\cite{FS}. As we shall see, due to non-commutativity such
examples can be found.

\section{Perturbative behavior of non-commutative Yang-Mills theories}
Recently, several authors\cite{KW,BS} analyzed the renormalization
behavior of non-commutative Yang-Mills theories (for an earlier
discussion see\cite{filk}. Related works are \cite{MS,SJ}). We briefly review
 their analysis. The Feynman
 rules of the commutative and non-commutative theories in momentum
 space are very
 similar. In fact the only difference is that 
 each vertex of the
non-commutative theory acquires a phase, $\exp{i \Sigma _{i<j} k_i
  \wedge k_j}$ (the Moyal phase), with respect to  
the vertex of ordinary commutative theory\cite{filk}. For planar diagrams, this
phase cancels at internal loops and the only
remnant is an overall phase. Therefore the planar commutative and
non-commutative theories are similar, in accordance with recent
findings\cite{HI,MR,LW}. In particular, thermodynamical quantities
 such as the free energy and the entropy are not affected by
non-commutativity at the planar limit \cite{MR}.

Another claim\cite{BS,KW} is that the oscillations of the Moyal phases at
high momentum would regulate non-planar
 diagrams, namely that UV divergences of non-planar diagrams
 would disappear. There are two exceptional cases in which non-planar
 diagrams  will still diverge\cite{KW}: (i). When the non-planar
 diagrams consist of
 planar sub-diagrams which might diverge. (ii). For specific values of
 momentum (a zero measure set) the Moyal phase can be zero. Indeed, 
 it was shown lately \cite{MSV} that in theories which contain scalars -
 there are non-planar divergences. The theories that we will 
 consider contain scalars and therefore contain infinities, however
 \cite{MSV} interpret these divergences as {\em Infra-Red}
 divergences. The reason is as follows: The integrals which are
associated with non-planar graphs converge unless the 
Moyal phase is set to zero by a vanishing incoming momentum. In this
case new type of divergences appear, and it seems that infinite number
of counterterms are needed. Therefore the non-commutative theory seems to be
 non-renormalizable. However, the authors of ref.\cite{MSV} suggest
that the re-summation of these divergent non-planar graphs would yield 
a finite result. Their observation is based on the similarity between
the present case and the standard (commutative) IR divergences.
Note that Infra-Red divergences are, anyways, expected in theories
 with massless particles. Thus, 'truly' divergences
 in the non-commutative theory occur only in planar
 graphs. 

\section{Finiteness of non-commutative orbifold field theories} 

Let us now consider an orbifold truncation of non-commutative \Nfour SYM.
These theories can be defined perturbatively by a set of Feynman
rules. The natural definition would be to attach to each vertex the
corresponding Moyal phase\cite{filk}. 
According to ref.\cite{BS}, the only potential divergences are the
ones which arise in planar diagrams. Non-planar diagrams are expected
to be finite, except for some ``accidental'' divergences \cite{BS}.
 Note that the infinities which do occur in these graphs
 are associated with the Infra-Red \cite{MSV} and therefore do not
contradict our claim that orbifold field theories are UV finite.

According to the analysis of \cite{BJ}, the planar diagrams
of an orbifold theory can be evaluated by using the corresponding diagrams of the parent
 non-commutative \Nfour. These diagrams are finite
 since they differ from the commutative
\Nfour only by an overall phase. In this way sub-divergences of non-planar
diagrams will also be canceled.
We therefore conclude that any orbifold truncation of non-commutative
non-compact \Nfour SYM is finite.
 In particular, it means that we might have a rich class of
non-supersymmetric gauge theories which are finite, even at finite 
$N_c$ (in contrast to ordinary non-supersymmetric orbifold field
 theories, where the two loop beta function is generically non-zero). 
Note that though these theories might be finite, they are certainly not 
conformal.

Let us consider a specific example\cite{KT}. The example is an
 $SU(N) \times SU(N)$ gauge theory with 6 scalars in
 the adjoint of each of the gauge groups
 and 4 Weyl fermions in the $(N,\bar N)$ and 4 Weyl fermions in
 the $(\bar N,N)$ bifundamental representations. 
It is the theory which lives on dyonic D3 branes
of type 0 string theory and can be also understood, from the
field theory point of view, as a $Z_2$ orbifold
 projection of \Nfour SYM\cite{NS}. The large $N_c$ commutative theory
 contains a line of fixed points. We suggest that its analogous
 non-commutative (finite $N_c$) theory is finite at that line. Namely, at
$g_1=g_2=h_1=h_2$, where $g_1,g_2$ are the gauge couplings and
$h_1,h_2$ are the Yukawa couplings. Note that in contrast to the
 commutative case the position of the line of finite theories
is {\em not} corrected by ${1\over N_c}$ contributions.

Finally, it might be interesting 
to understand how the finiteness of these theories arise from string
 theory orbifolds.

\Acknowledgements
I would like to thank O. Aharony, N. Itzhaki and B. Kol for useful
 discussions and comments. This work was supported in part by EEC TMR
contract ERBFMRX-CT96-0090.

\end{document}